# Impact of Mobile phone in the Air and Random Access Channel (RACH) with Time Division Multiple Access (TDMA) Noise in Aircraft Avionics

**Lau Siow Shyong**

**lausiowshyong@gmail.com Graduated from Texas A&M University (College Station)**

**Worked for TRL Wireless, Goldtron, Motorola and iwow before joining the faculty of Aerospace Electronics at Temasek Polytechnic  lauss@tp.edu.sg**

## ABSTRACT

Mobile phone already allowed to be used in some commercial airlines and the first standard to be deployed is 1800MHz GSM band. After Many studies and debate, the regulation authorities have spelled out the requirement for picocell and mobile phone to be used in aircraft so as not to be interfered with avionics system in the air. The main objective is to limit the mobile phone at the lowest transmitting power controlled by the picocell installed in the aircraft, at the same time, to prevent the mobile phone from trying to camp on to the Ground Base Station.

There are quite a number of different frequency bands and different technologies implemented in today's mobile phone. 1800MHz GSM band was selected to be the first technology used in air because it is simpler and easier to control. Many studies talked about lower the mobile phone output power to the lowest of 0dBm, but seldom talk about the Random Access Channel (RACH), which will emit at the highest power of 30dBm at the instant of call making. This will occur at "time slot 0" and is not controllable until the connection between the Base Station and mobile is made.

We will take about the impact of RACH and hence the TDMA noise generated in the aircraft in more detail.

## 1. Introduction

Today, most of us could not leave home without mobile phone. How about in the aircraft?

The pressure of airlines to provide mobile phone service is getting higher and higher. And after many studies and debate, the aviation authorities decided to give the green light for mobile phone usage in the air.

The first technology to be deployed is GSM 1800MHz band. Right now, we have 2G (GSM), 3G (WCDMA), 4G (LTE) RF capabilities in our smart phone. GSM 1800 MHz is relative stable and simpler, could be controlled easily by the picocell deployed in the aircraft.

Picocell acts like a Base station in the air and controls all the mobile phones within the range. The picocell will command the mobile phone to transmit at the lowest power level in the air. For 1800MHz GSM, it will be power level 15 which is 0dBm.

Fig 1 shows the power level chart of GSM1800MHz.

|  | DCS 1800MHz |  |  |
|---|---|---|---|
| Output Power | Normal burst Ch513 to Ch884 |  |  |
| Normal Burst | TX0 (30 dBm) | 30 | dBm |
|  | TX1 (28 dBm) | 28 | dBm |
|  | TX2 (26 dBm) | 26 | dBm |
|  | TX3 (24 dBm) | 24 | dBm |
|  | TX4 (22 dBm) | 22 | dBm |
|  | TX5 (20 dBm) | 20 | dBm |
|  | TX6 (18 dBm) | 18 | dBm |
|  | TX7 (16 dBm) | 16 | dBm |
|  | TX8 (14 dBm) | 14 | dBm |
|  | TX9 (12 dBm) | 12 | dBm |
|  | TX10 (10 dBm) | 10 | dBm |
|  | TX11 (8 dBm) | 8 | dBm |
|  | TX12 (6 dBm) | 6 | dBm |
|  | TX13 (4 dBm) | 4 | dBm |
|  | TX14 (2 dBm) | 2 | dBm |
|  | TX15 (0 dBm) | 0 | dBm |

**Fig 1. Transmit power level for GSM1800MHz**

On the ground, the Base Station commands the mobile phone transmits' power level depending on the signal strength the phone is receiving. So as to ensure good signal quality and battery life saving.

In the air, the mobile phone could only transmit at the lowest power (0dBm) to avoid any interference to the avionics system in the aircraft.

The other technical challenge is to ensure that the mobile phone in the air will not attempt to camp on to the base station on the ground. This is done by introducing a Noise floor by the Network Control Unit (NCU) on the plane. The NCU Noise floor level is set at 12 dB lower than the GSM signal since the typical Signal to Noise ratio of GSM is 9dB, plus 3dB margin of all others interference.

So, we have to take care of the avionics systems in the aircraft by commanding the mobile phone to transmit at the lowest power level, and preventing the mobile phone from attempting to camp on to the Ground Base Station.

But, seldom we talked about the Random Access Channel (RACH), which will emit at the highest power of 30dBm at the instant of call making. This will occur at "time slot 0" and is not controllable until the connection between the station and mobile is made.

We will take about the impact of RACH and hence the TDMA noise generated in the aircraft in more detail.

## 2. Mobile phone radiated power to the Avionics system

3GPP organization has spelled out the radiated spurious limitation of mobile phone. But in this case, we would be more interested in the carrier frequency and it's harmonics of the mobile phone that could be possibly interfered to the avionics systems.

So, the frequency is 1800MHz and the multiply of it. Power level has been lowered to 0dBm by the picocell in the aircraft. We will talk about RACH later.

## 2.1 Avionics Frequencies

Fig 2 shows the Main Avionics Equipment Frequency Spectrum.

**Main Avionics Equipment Frequency Band**

| | |
|---|---|
| HF communication systems | 2MHz to 30MHz |
| VHF communication systems | 118 MHz to 136.975 MHz. |
| Marker Beacon | 75MHz |
| VOR | 108.00 to 117.95MHz |
| Localizer | 108.10-111.95 |
| Glideslope | 329.15-335 MHz |
| DME | 962 to 1213MHz |
| GPS | 1.575GHz |
| Satellite at L-band frequencies | 1,530 MHz to 1,660.5MHz |
| Doppler navigation | X-band (8.8-9.8 GHz) |
| Weather radar | C Band (4 - 8GHz) |
| Radio Altimeter | 4,250MHz and 4,350MHz |

**Fig 2 Main Avionics Frequency Band**

From **Fig 2,** the DME is operating from 962MHz, and that is close to the GSM900MHz band edge of 960MHz. But for GSM1800MHz, there is no direct interference except that we need to be careful for the 3<sup>rd</sup> harmonics of 1800MHz which is close to the Weather radar.

## 2.2 Interference Calculation and Measurement

We could start with theoretical calculation and verify with measurement result.

The interference source will be the GSM1800 transmitter output power and its harmonics. There are some other interference sources like the 13MHz or 26MHz Reference Clock, although it's harmonics could go quite far up to 1GHz, the amplitude should be low. The other intermodulation products are not high as well.

We could assume 0dBi Antenna gain for mobile phone. Minus near field gain, with propagation loss, roughly we could calculate the Field strength over the distance. Or we could get the Radiated limitation template from 3PGG as a reference, every phone supposed to compliance to this specification.

The victim will be the Avionics equipment receiver. Get the sensitivity and selectivity of the Avionics equipment. The Co-channel, Adjacent channel and Blocking specification.

For this case, there is no direct hit of frequency. Hence, just by looking at the blocking specification will give us an ideal whether the Avionics' Receiver sensitivity be degraded by the interference signals.

To make practical measurement, ensure that we could read the sensitivity of the Avionics equipment first, turn on the mobile phone at the maximum usable power to see any degradation of sensitivity.

## 3. Random Access Channel (RACH) impact

The first technology to be granted for use in aircraft is GSM1800MHz. Picocell deployed in the aircraft will command the mobile phone to be transmitted at lowest power level of 0dBm. There are many studies and measurement being made for the safety use of mobile phone in the aircraft.

But not much discussion about the RACH impact on the Avionics equipment.

Every time we turn on our mobile phone, it will search for the signals that belonged to it and camp on to the station. When the mobile want to make a call or an initial contact, A RACH will be sent to the base station at the highest power level of the phone. In this case, is 30dBm (Ref. to Fig 1). This will occur at "time slot 0" and is not controllable until the connection between the station and mobile is made.

This is the protocol of GSM defined in 3GPP. It is resided in the phone once it is manufactured. Unless the Picocell in the aircraft could change it in the air, we will have this instant highest power emitted from the mobile phone when it is trying to make a call.

# 4. Random Access Channel (RACH) Time Slot and Frame

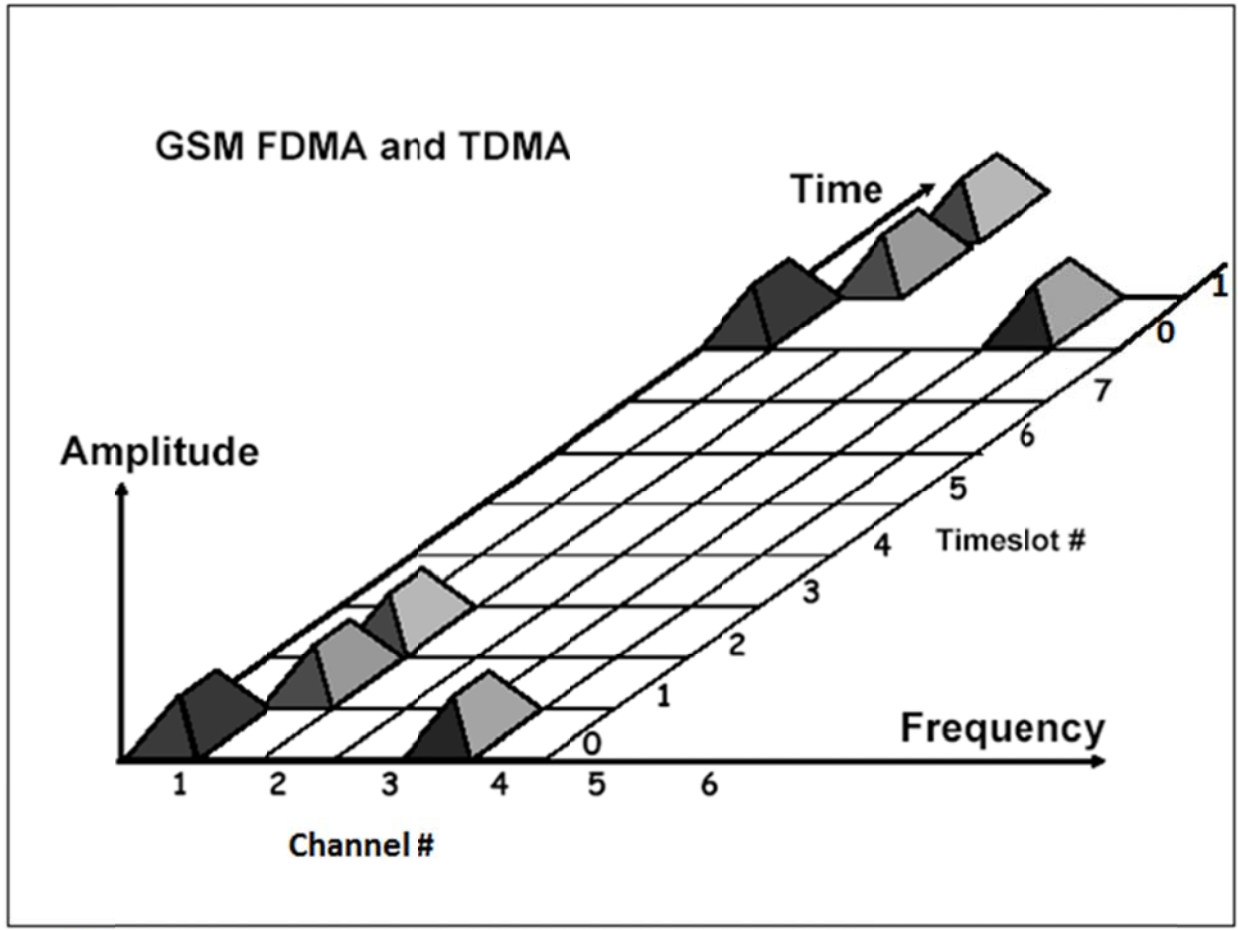

**Fig 3. GSM Frequency Division and Time Devision**

Refer to **Fig 3**. GSM is Frequency Division Multiple Access (FDMA) and Time Division Multiple Access (TDMA). Each frequency channel is divided into 8 time slots. The actual data transfer occurs from time slot 1 to time slot 7 (Traffic Channel). Time slot 0 is reserved for system controlling (Control channel). RACH belonged to Control Channel and only occur at time slot 0. Refer to **Fig 4** for the time frame reserve for RACH.

### BCH & CCCH - 51 Frame Structure - UpLink

**Frame number ( Up Link )**

| Time Slots | 0 | 1 | 2 | 3 | 4 | 5 | 6 | 7 | 8 | 9 | 10 | 11 | 12 | 13 | 14 | 15 | 16 | 17 | 18 | 19 | 20 | 21 | 22 | 23 | 24 | 25 |
|---|---|---|---|---|---|---|---|---|---|---|---|---|---|---|---|---|---|---|---|---|---|---|---|---|---|---|
| 0 | D | D | D | D | R | R | H | H | H | H | H | H | H | H | R | R | R | R | R | R | R | R | R | R | R | R |
| 1 | | | | | | | | | | | | | | | | | | | | | | | | | | |
| 2 | | | | | | | | | | | | | | | | | | | | | | | | | | |
| 3 | | | | | | | | | | | | | | | | | | | | | | | | | | |

| | 26 | 27 | 28 | 29 | 30 | 31 | 32 | 33 | 34 | 35 | 36 | 37 | 38 | 39 | 40 | 41 | 42 | 43 | 44 | 45 | 46 | 47 | 48 | 49 | 50 |
|---|---|---|---|---|---|---|---|---|---|---|---|---|---|---|---|---|---|---|---|---|---|---|---|---|---|
| 0 | R | R | R | R | R | R | R | R | R | R | R | D | D | D | D | D | D | R | R | D | R | D | D | D | D |
| 1 | | | | | | | | | | | | | | | | | | | | | | | | | |
| 2 | | | | | | | | | | | | | | | | | | | | | | | | | |
| 3 | | | | | | | | | | | | | | | | | | | | | | | | | |

SDCCH ( D ) : Standalone Dedicated Control Channel
SACCH ( H ) : Slow Associated Control Channel
PCH ( P )　　: Paging Channel
RACH ( R )　 : Random Access Channel

**Fig 4. GSM Frame Structure**

RACH will only occupy less than a time slot, it is a highest power burst during call set up. It could transmit several time before connection is successfully made and before it could be commanded to transmit at lower power level. Hence, beside considering lowest power level of GSM1800 of 0dBm. We need to investigate the highest power level of 30dBm RACH burst.

## 5. Multiple Transmission

GSM is FDMA and TDMA system where there is no sharing of traffic channel. Everyone will get the dedicated physical line for communication. One exception is the RACH. Although there are some dedicated time frames for the RACH transmission, multiple phones might send RACH signal at the same time at the highest power level since Base Station has no full control of the mobile phone yet

In fact, besides making call, RACH could happen in some other situation:

• making emergency call

• making multiple call re-establishment following link failure

• answering to message paging

• making location updating procedure.

So, there is a possibility that multiple passengers in the airplane send RACH signals at the same time. This will make the scenario slightly more complicated.

Due to Fading, multipath and phase different, two signals might not just add together. But we could still study the worst case scenario assuming some of the signals will be added together.

By looking at the blocking specification of the Avionics equipment, predict the maximum signals strength it could stand. Based on Root Sum Square (RSS) and Antenna Path Loss behavior, get the distance required for the mobile phone to be safely used.

## 6. Time Division Multiple Access (TDMA) Noise

You may experience this when your mobile phone is near a speaker, an audible "Buzz" sound. This is due to the high RF energy being coupled into the Speaker audio path. It rectifies the RF signal and extracts the TDMA envelope in the way an AM detector would. It is more noticeable at the instant when you make a call because that is when the RACH highest power sent. After the physical connection is established, the mobile phone will be transmitting at lower power level based on the command by the Base Station.

Refer to **Fig 5** GSM frame. The Transmitter will emit one time slot per frame. This periodic signal will create a 217Hz audible signal and hence the "Buzz" sound.

lausiowshyong@gmail.com

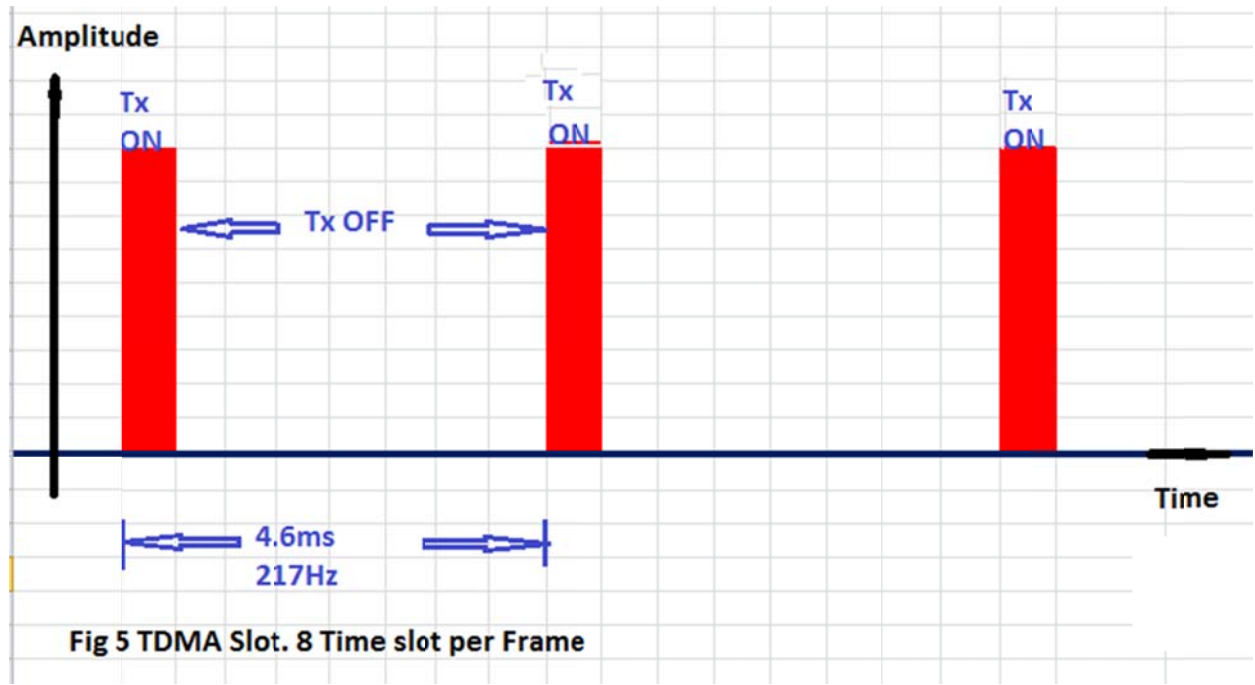

Fig 5 TDMA Slot. 8 Time slot per Frame

This is a typical technical problem for mobile phone designer since the phone is getting smaller and smaller, the speaker in the phone is placed near to Antenna, Power Amplifier or Battery. Two main reasons for the coupling: DC power supply line being dipped by the high current drawn; RF radiated power.

## 6.1 Minimize the "Buzz" sound

Power Amplifier draws high current when transmitting at the Tx time slot, this create a dip in the supply line and ripple in the Ground. This will propagate to other audio circuitry and speaker and microphone. Refer to **Fig 6** for the typical dipped supply line.

Quick methods to minimize the TDMA noise for this case are:

- Adding large capacitors on the supply line to reduce the dip.
- Adding frequency dependent smaller capacitors to couple away the carrier frequency. Practical result showed that 18 pF is good for this use. The self-resonating frequency is around 1800Mz. Please refer to **Fig 7.** for the implementation.
- Power supply line to be separated with the Audio path.
- Separate Ground for the Audio and Supply as well as the RF path.

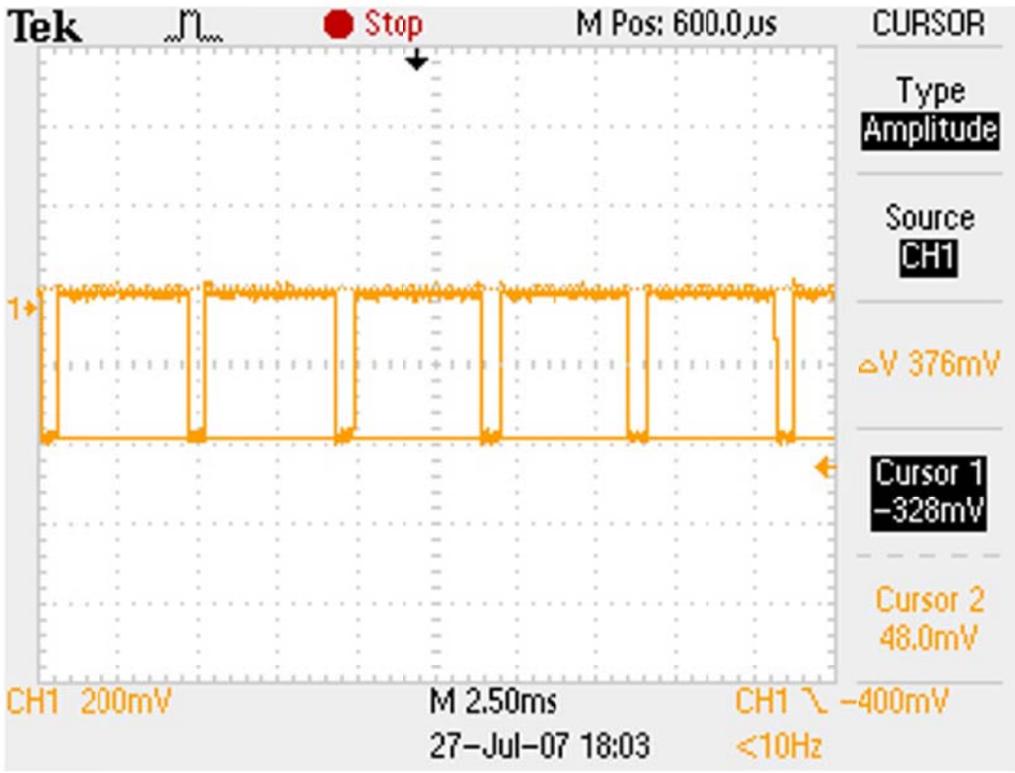

Fig 6. A typical supply line dip due to TDMA burst

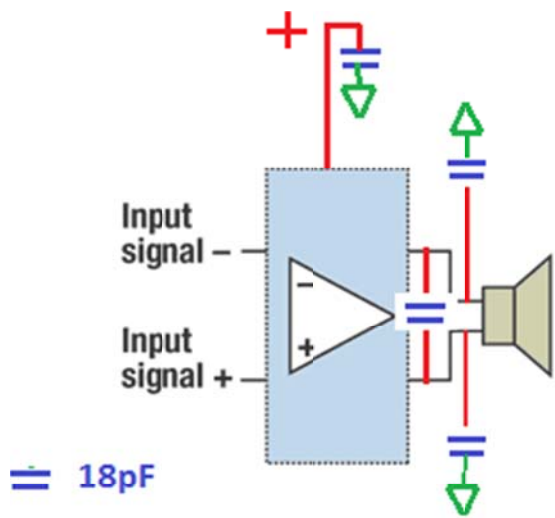

Fig 7. Coupling Capacitor on Audio Path

lausiowshyong@gmail.com

### 6.2 Minimize the "Buzz" sound in the aircraft

The "Buzz" sound may happen in the aircraft where the speakers in the cabin pick up the RF energy from the mobile phone especially when the RACH is transmitting. Although it will not cause any safety issue, but might sound alarming.

The approach to minimize the "Buzz" may be the same as what the mobile phone designer normally do as stated above. The good thing is that it is not confined to a smaller area as the mobile phone. But the complex area will be the Ground issue and the long wire that is running across the cabin.

The ultimate aim is to reduce the AM radiated noise to the speaker.

## 7. Conclusions

Many studies have shown that it is safe to use GSM1800 technology in the aircraft by commanding the mobile to transmit at the lowest power of 0dBm. But we should take precaution of the RACH signal with the highest power of 30dBm at the instant of calling. Especially when multiple RACH signal being sent by multiple users at the same time.

Beside the impact of Avionics equipment, TDMA noise might be heard in the cabin. The approach of minimizing the noise will be quite the same as what the mobile phone designer do.